\documentstyle[prd,aps]{revtex}
\def\RE {I\kern-6pt R    }
\def\Z  {Z\kern-13pt Z   }
\def\be {\begin{equation}}
\def\ee {\end{equation}  }
\def\beq{\begin{eqnarray}}
\def\eeq{\end{eqnarray}  }

\def\bi {\begin{itemize} }

\def\ei {\end{itemize}   }
\def\gtwid{\mathrel{\raise.3ex\hbox{$>$\kern-.75em\lower1ex\hbox{$\sim$}}}}
\def\ltwid{\mathrel{\raise.3ex\hbox{$<$\kern-.75em\lower1ex\hbox{$\sim$}}}}

\input epsf

\begin{document}

\title{Critical behavior of gravitating sphalerons}

\author{R. Steven Millward\footnote {Electronic address:  
                                       \tt rsm52@email.byu.edu\hfil} 
   and  Eric W. Hirschmann\footnote {Electronic address:  
                                       \tt ehirsch@kepler.byu.edu\hfil}} 
\address{Department of Physics and Astronomy,
         Brigham Young University,
         Provo, Utah 84604}

\maketitle

\begin{abstract}
We examine the gravitational collapse of sphaleron type configurations 
in Einstein--Yang--Mills--Higgs theory.  Working in spherical symmetry, we 
investigate the critical behavior in this model.  We provide
evidence 
that for various initial configurations, there can be three different 
critical transitions between possible endstates with different critical 
solutions sitting on the threshold between these outcomes.  In addition,
we show that within the dispersive and black hole regimes, there are 
new possible endstates, namely a stable, regular sphaleron and a stable, 
hairy black hole.  
\end{abstract}

\section{Introduction}
\label{sec:introduction}

Over the last two decades, substantial effort has been brought to bear
on the study of gravitating Yang--Mills fields.  This research has resulted 
in the discovery of numerous solutions to the coupled 
Einstein--Yang--Mills(--Higgs and/or --dilaton) equations.  These solutions
include both black hole and regular or particle--like solutions.  In addition
to confirming the richness of these nonlinear systems, this work has 
also been helpful in clarifying the standing of the black hole uniqueness 
theorems and various ``no-hair" ideas.

While many of these solutions have been found by solving the appropriate static 
equations, it was realized early on that understanding the 
stability of these solutions is important in order to ascribe relative 
significance to these solutions within the context of some of the no--hair
conjectures.  The primary means for evaluating stability have been linear 
perturbation analyses of various static solutions.  As a result, many of these 
static, gravitating Yang--Mills(--scalar) configurations
have been found to be unstable to small time--dependent perturbations.  

That many of these solutions appear unstable in linear perturbation theory
does not necessarily mean that such solutions are without significance.  
Indeed, it is now widely accepted within the context of gravitational 
critical phenomena that some of these solutions will have relevance as
attractors in the critical collapse of gravitating fields at the threshold
of black hole formation \cite{Chop,Gund}.  As an example, the Bartnik--McKinnon solutions of 
the spherically symmetric, static Einstein--Yang--Mills equations are a 
countably infinite family of regular solutions characterized by the 
integer number, $n$, of 
zero--crossings of the gauge potential.  These solutions are unstable in
linear perturbation theory with the $n^{\rm th}$ member of the family having
$n$ unstable modes.\footnote{Strictly, this is true if only the radial,
gravitational perturbations are excited.  If additional components of the 
gauge potential are also perturbed ({\it i.e.} the sphaleron sector), 
the $n^{\rm th}$ solution will have $2n$ unstable modes, the sphaleron sector
contributing an additional $n$ unstable modes to those from the gravitational
sector.\cite{LavMai}}  Thus, the $n=1$ member of this family has a single 
unstable mode
and is a candidate for being a critical solution in the gravitational
collapse of a pure Yang--Mills field.  Indeed, this is exactly the result 
found in \cite{Chchbi} where the full evolution equations for the model
were solved.  The $n=1$ Bartnik--McKinnon solution is the critical solution
that sits on the threshold between the complete dispersal of the collapsing field 
and the formation of a finite size black hole (Type I collapse).  The
value of the unstable mode for this solution then correctly predicts the 
scaling relation for the lifetime of near--critical solutions. 

In addition to these regular solutions, the Einstein--Yang--Mills equations 
also admit another countably
infinite family of solutions, but with a nonzero horizon.  These black hole
solutions have non-trivial hair outside their horizons and are again 
characterized by the number, $n_{\rm BH}$, of zero-crossings of the gauge 
potential.  They too are unstable in linear perturbation theory 
with $n_{\rm BH}$ unstable modes from the gravitational sector.  In agreement 
with expectations, it has been shown that the $n_{\rm BH}=1$ solution is 
also a critical solution \cite{Chhima}.  But this non-abelian or colored 
black hole solution, rather than separating dispersion and black hole 
formation as does the $n=1$ Bartnik--McKinnon solution, sits on the 
threshold between two different kinds of dynamical collapse.

Given this behavior, it is natural to conjecture that other configurations
of gravitating Yang--Mills fields should likewise exhibit critical phenomena. 
With that in mind, we consider here the nonlinear evolution of gravitating 
$SU(2)$ sphalerons in Einstein--Yang--Mills--Higgs theory.  We provide 
evidence  
that there can be three critical transitions in the initial data space.  
These include the now standard Type I and Type II transitions as well as 
the transition mentioned above between different kinds of dynamical 
collapse on which sits a colored or ``hairy" black hole as the 
intermediate attractor.  In the process of examining critical gravitational 
collapse within this system and the formation of such ``hairy" black holes as 
attractors in the black hole regime, we have also confirmed the stability  
of two additional endstates of collapse.  One is a regular, gravitationally 
bound configuration of Yang--Mills--Higgs field forming a stable 
``sphaleron star."  The second is a family of stable, hairy black
holes different from those that serve as the critical solutions in the black
hole regime.  The existence of such stable solutions appears to have first been
predicted by Maison \cite{Maison}.   

The outline of the remainder of the paper is then as follows.  Section 2 
summarizes the equations which constitute the full evolution problem 
and our numerical approach to their solution.  Section 3 describes the 
numerical results, including results of our parameter space searches, 
the critical solutions and the nature of the 
stable solutions.  Section 4 offers some conclusions and thoughts for future 
directions.

\section{The model}
\label{sec:model}

Our starting point in studying the gravitational collapse of configurations of 
Yang--Mills--Higgs fields is the action
\beq
S & = & \int d^4x \sqrt{-g} \left[   {1\over {16{\pi}G}}R 
                                   - {1 \over 4g^2} {F^a}_{\mu \nu} F^{a\mu \nu}
                                   - {1 \over 2} ({D_{\mu} \Phi})^{\dagger} 
                                                   D^{\mu} \Phi 
                                   - V(| \Phi |^2)  
                           \right]  
\eeq
where $F^a_{\mu\nu}$ is the Yang-Mills field strength tensor 
given by 
\beq
F^{a}_{\mu\nu} & = &   \partial_{\mu} A^{a}_\nu 
                     - \partial_{\nu} A^{a}_{\mu} 
                     + \epsilon^{abc} A^{b}_{\mu} A^{c}_{\nu} , 
\eeq
$D_\mu$ is the gauge covariant derivative whose action on the Higgs
doublet, $\Phi$, is
\beq
D_{\mu}\Phi = \nabla_{\mu}\Phi + A_{\mu} \Phi ,  
\eeq
and the potential $V$ is taken to be 
\beq
V( |\Phi|^2) & = & {\lambda \over 4} \, \big( \Phi^\dagger \Phi - \eta^2 )^2.
\eeq
Varying the action with respect to the metric, the field strength and the Higgs
field result in the Einstein equations and the curved space Yang-Mills 
and Higgs equations, respectively.  These are 
\beq
{1\over 8\pi G} G_{\mu\nu} & = &  {1 \over g^2} 
           \left[   F^{a}_{\mu\lambda} F^a{}_\nu{}^\lambda
                 - {1\over4} g_{\mu\nu} F^{a}_{\alpha\beta} F^{a\alpha\beta}
          \right] 
\nonumber\\
&  &  \qquad + \left( D_{\mu}\Phi \right)^{\dagger} D_{\nu}\Phi
             - {1\over2} g_{\mu\nu} 
                    \left(D_{\lambda}\Phi \right)^{\dagger} D^{\lambda}\Phi
             - g_{\mu\nu} V( |\Phi|^{2}) 
\\
D_{\mu} F^{a\mu\nu} & \equiv & \nabla_{\mu} F^{a\mu\nu} + \epsilon^{abc} A_{\mu}^{b}
F^{c\mu\nu} = 0 .
\\
D^{\mu} D_{\mu} \Phi & = & \lambda \, \Phi (\Phi^{\dagger}\Phi - \eta^2)
\eeq

With these general forms for the 
equations of motion, we make some simplifying assumptions.
In particular, we will restrict ourselves to spherically symmetric 
gravitational collapse and 
work exclusively with an $SU(2)$ gauge group.  We also make the assumption 
that the Higgs field lives in the fundamental representation of $SU(2)$.  The
corresponding flat space version of this theory includes 
the so--called sphaleron solutions \cite{Dahane}.  

Our intent is to solve the full set of nonlinear, evolution equations 
representing gravitational collapse.  In order to do this numerically we must 
fix both the coordinate freedom and the gauge freedom in our model.  There
are, of course, numerous possibilities, but we will try to hew fairly closely 
to related work of others in the field.  For the coordinate system, we will 
work in maximal, areal coordinates.  If the general form of the spherically 
symmetric metric is written as  
\beq
ds^2 & = & (- {\alpha}^2 + a^2 {\beta}^2) dt^2 + 2 a^2 \beta dt dr + a^2 dr^2
+b^2 r^2 {d \theta}^2 + b^2 r^2 \sin^2 \theta {d \phi}^2  
\eeq
where the metric components depend only on $t$ and $r$, the choice of 
areal coordinates amounts to $b=1$ while choosing maximal time slices 
corresponds to the vanishing of the trace of the extrinsic curvature, $K=0$.  

For the Yang-Mills field, the most general form for a spherically symmetric
gauge potential is the Witten ansatz \cite{Wi}: 
\beq
A & = &   u {\tau}_r  dt 
        + v {\tau}_r dr 
        + \left( w{\tau}_{\theta} + ({\tilde w}-1){\tau}_{\phi} \right) d\theta
        + \left(   (1- {\tilde w}) {\tau}_{\theta} 
                 + w {\tau}_{\phi}
          \right) \sin\theta d\phi  
\eeq
where the $\tau_i \,$ ($i\in \{r,\theta,\phi \}$) are the spherical 
projection of the Pauli spin matrices and form an anti-Hermitian basis 
for the group $SU(2)$ satisfying $[\tau_i , \tau_j] = \epsilon_{ijk} \tau_k$.  
With this ansatz for the gauge potential there is some gauge freedom that 
allows us to simplify its form, namely the potential is invariant under 
a transformation of the form $U=e^{\psi(t,r) \tau_r}$.  We 
can fix some of that gauge freedom by choosing $v\equiv0$.  This choice 
effectively eliminates the $r$ dependance in the above gauge transformation.  
If we choose to work within the so-called ``magnetic ansatz" we can fix the 
remaining freedom in the following way.  It can be shown that in this 
ansatz, the component $u$ is a function only of $t$, {\it i.e.} it is now 
pure gauge and can be set to zero as part of our gauge fixing.  The 
remaining fields, $w$ and $\tilde w$, under the remaining constant gauge 
transformations are merely sent into linear combinations of each other 
and hence we can fix the last bit of gauge freedom by setting $w=0$.  
This leaves  $\tilde{w}$ as the sole non-zero component of the gauge 
potential.

Our form for the Higgs field, taken from \cite{Grmaon}, is  
\beq
\Phi & = & {1 \over \sqrt 2} ( \gamma {\bf 1} - 2 \psi \, \tau_r ) 
            \pmatrix{0\cr 1}  
\eeq  
and though not strictly spherically symmetric, results in a spherically 
symmetric energy density \cite{Maison}.  We will consider in this work 
only the case in which  
$\psi = 0$.  This is not a gauge choice but an additional assumption 
made merely to simplify the
resulting equations and dynamics.  A similar thing is done, for instance, 
in \cite{Grmaon}.  

With these assumptions, the evolution equations for the
Yang-Mills field become 
\beq
\dot{\tilde w} & = & {\alpha \over a } \, P + \beta Q  
\\  
\dot{Q}        & = & \biggl({\alpha \over a} P + \beta Q \biggr)'  
\\
\dot{P}        & = &    \biggl( \beta P + {\alpha \over a} Q \biggr)' 
                      + {\alpha a \tilde{w} \over r^2} \, 
                        \bigl(1 - {\tilde{w}}^2 \bigr) 
                      - {\alpha a \over 4 } \, \gamma^2 \, 
                        \bigl(\tilde{w} - 1\bigr)  
\eeq
while the evolution equations for the Higgs field are given as 
\beq
\dot {\gamma} & = & {\alpha \over a} \, G + \beta E  
\\
\dot{E} & = & \biggl({\alpha \over a} G + \beta E\biggr)'  
\\
\dot{G} & = &   {1\over r^2} \biggl[r^2 (\beta G + {\alpha \over a} E)\biggr]'
              - { \alpha a \gamma \over 2 r^2} \, \bigl(\tilde{w} - 1\bigr)^2 
              - {\lambda \over 2} \, 
                \alpha a \gamma \bigl({\gamma}^2 - 2 {\eta}^2\bigr) ,
\eeq
where, as usual, overdots and primes denote differentiation with respect to 
$t$ and $r$ respectively.  
Both of these sets of evolution equations are supplemented with the 
first order definitions 
$Q = \tilde{w}'$ and $E = \gamma'$ as well as the constraints on the metric
components coming from the Einstein equations   
\beq
\alpha'' & = &   \alpha' \biggl({a' \over a} - {2 \over r}\biggr) 
               + {2 \alpha \over r^2} \biggl( a^2 -1 + {2ra' \over a} \biggr) 
               + 4 \pi G \alpha ( S - 3 \rho )  
\\ 
a' & = &   a \, { 1-a^2 \over 2r} 
         + {3 \over 8} r a^3 K_{r}{}^{r}{}^2 
         + 4 \pi G \, r a^3 \rho.  
\\
K_{r}{}^{r}{}' & = & - {3 \over r} K_{r}{}^{r} 
                     + 8 \pi G \, {1\over a} \Bigl[   {PQ \over r^2} 
                                                    + {1 \over 4} EG \Bigr] 
\\
\beta & = & \alpha r {K_{\theta}}^{\theta} 
\eeq
The matter stress-energy terms in these equations are given by 
\beq
\rho + S_r{}^r & = &   {P^2 + Q^2 \over a^2 r^2}
                     + {E^2 + G^2 \over 4 a^2}
\\
\rho - {S_r}^r & = &   { ({\tilde{w}}^2 -1)^2 \over 2 r^4 }
                     + { \gamma^2 \, (\tilde{w}-1)^2  \over 8 r^2 }
                     + {\lambda \over 16} (\gamma^2 - 2 \eta^2)^2
\\
{S_{\theta}}^{\theta} & = & {1 \over 8 a^2}(G^2 - E^2) + {{({\tilde{w}}^2 -1)}^2
\over 4 r^4} - {\lambda \over 32}{({\gamma}^2 - 2 {\eta}^2)}^2
\\
j_r & = & -{PQ \over a r^2} - {EG \over 4 a}
\eeq

Boundary conditions are implemented by demanding regularity at the origin and 
requiring the presence of only outgoing radiation at large distances (see 
\cite{Chhima}).  The resulting constraints on the metric components require 
$\alpha'(t, 0)=a'(t, 0)=\beta(t, 0)=K_{r}^{r}(t, 0)=0$.  The matter fields may 
satisfy one of two possible regular configurations at the origin.  Either, 
$\gamma(t, 0)=0$ and $\tilde{w}(t, 0)=-1$, or $\gamma'(t, 0)=0$ and 
$\tilde{w}(t, 0)=1$.  These two choices correspond to the odd and even 
node solutions, respectively \cite{Grmaon}.  In order to find the
critical solution we choose the former and look for the solution with a single
unstable mode.  The outgoing conditions require that at the edge of our grid, 
\beq
P & = & -Q  \\ 
\dot{Q} & = & [({\alpha \over a} - \beta)Q]'
\\
G & = & {1 \over r}(\beta{a \over \alpha}-1)(\gamma - \sqrt{2}\eta)
\\
\dot{E} & = & [(\beta -{\alpha \over a})(E + {1 \over r}[\gamma - \sqrt{2}\eta])]'.
\eeq

 For the initial pulse, we use a ``time-symmetric kink" as in \cite{Chhima} 
for the gauge potential, namely
\beq 
w(0,r) & =  & \left[ 1 + a \left(1 + {br \over s}\right)
            e^{-2\left(r/s\right)^2} \right]
                  \cdot \tanh \left( {x-r \over s} \right) \\ 
\dot{w}(0,r) & =  &  0 
\eeq
where the parameters $a$ and $b$ are chosen so $w(0, 0)=-1$ and $w'(0, 0)=0$.  
The parameters $x$ and $s$ are the center and width of the kink, 
respectively.

The Higgs field is initialized as
\beq
\gamma(0,r) & = &    \sqrt{2} \, \eta \, \tanh{\frac{r}{\delta}}
                   + A_H \, e^{ -(r-r_0)^2/ d^2 } 
\\
\dot{\gamma}(0,r) & = & 0 
\eeq
where the parameter $\delta$ is usually set to $\delta=10$.  This is 
primarily due to the fact that varying $\delta$ does not significantly change  
the final result of the collapse.  As a consequence, we perturb the Higgs
field via a gaussian pulse.   
Similar to the initialization for the Yang--Mills field, the 
parameters $A_H$, $r_0$, and $d$ which describe the initialization of the Higgs
field represent the amplitude, center,
and width of the gaussian pulse, respectively.  These initial data parameters
for the Yang--Mills and Higgs fields will constitute our initial data set 
and will be used when tuning our evolutions to the critical solutions.

Our numerical approach closely follows that of \cite{Chhima}.  We use a uniform
grid recognizing that we will not have sufficient resolution to investigate
Type II collapse in a completely satisfactory way.  Nonetheless, we have 
indications verifying the existence
of Type II behavior in our model.  For this paper, therefore, we focus our 
primary interest on the black hole
transition and the dynamics occuring within the black hole regime.   

We use an iterative Crank--Nicholson scheme for the evolution equations while
for the constraints, we simply integrate outward from the origin.   
As we want to consider evolutions that extend to the future of black hole
formation, our use of maximal slicing is crucial.  In our coordinates, the 
apparent horizon equation is an algebraic relation  
\beq
a r {K^{\theta}}_{\theta} & = & 1
\eeq 
We use the same black hole excision technique developed in \cite{MaCh} and 
used in \cite{Chhima}.
As discussed there, we set a threshold value slightly larger than one such that
if $a r {K^{\theta}}_{\theta}$ exceeds that value for certain grid points, we 
discard those points at future time steps considering them inside the apparent
horizon.  At the boundary of this region, we need no new boundary conditions 
for the evolution equations.  For those variables solved via constraint 
equations we either switch to solving an evolution equation subsequent to the 
formation of a horizon or we ``freeze" the variable ({\it e.g.} $\alpha$) 
such that it retains the value it had when the horizon formed \cite{Chhima}.  
As a result, though we can observe matter falling into the horizon, 
we cannot comment on any dynamics within the apparent horizon as the  
evolution is effectively frozen for values of the radial coordinate less than
the horizon radius.    
This procedure thus allows us to evolve past the formation of the black 
hole and thereby investigate such things as the final endstates as well as the 
critical dynamics in the vicinity of transition regions.  

We have tested the resulting code and shown it to be second--order convergent
and to conserve mass.  It also reproduces the results of \cite{Chhima} in the 
limit that the Higgs field and its coupling vanish.  Finally, we note that 
we made extensive use of RNPL 
(Rapid Numerical Prototyping Language) \cite{RNPL}, a language written 
expressly to aid the differencing and solution of PDE's.

\section{Results}
\label{sec:results}

In attempting to evolve these equations, it quickly became clear that the
size of the initial data sets that could be varied is somewhat unwieldy 
and we had to make choices in order to restrict the possible  
sets of initial data parameters.  Although we have performed numerous 
evolutions by varying the elements of different sets 
of initial data parameters, we will focus on the evolution of two sets of 
parameters to highlight our results.  Other sets would appear to 
give qualitatively similar conclusions.  
When we evolve these equations, we confirm many of the same aspects that 
have come to be expected in similar models.  However, there are, at the same
time, a number of unexpected surprises.

To begin our examination of the dynamics of this model, we consider varying
two parameters describing the initialization of the gauge potential, 
namely $x$ and $s$, the center and width of the kink, respectively.  The 
amplitude of the perturbing, gaussian pulse for the Higgs field is set to 
zero, $A_H=0$, and the width of the $\tanh$ function describing the Higgs is
$\delta=10$.  We call this configuration the ``bland" Higgs field.  Note
that in this section, all the pictured results are for values of the Higgs
coupling parameters, $\eta = 0.1$ and $\lambda = 0.5$.   

\begin{figure}
\epsfxsize=15cm
\centerline{\epsffile{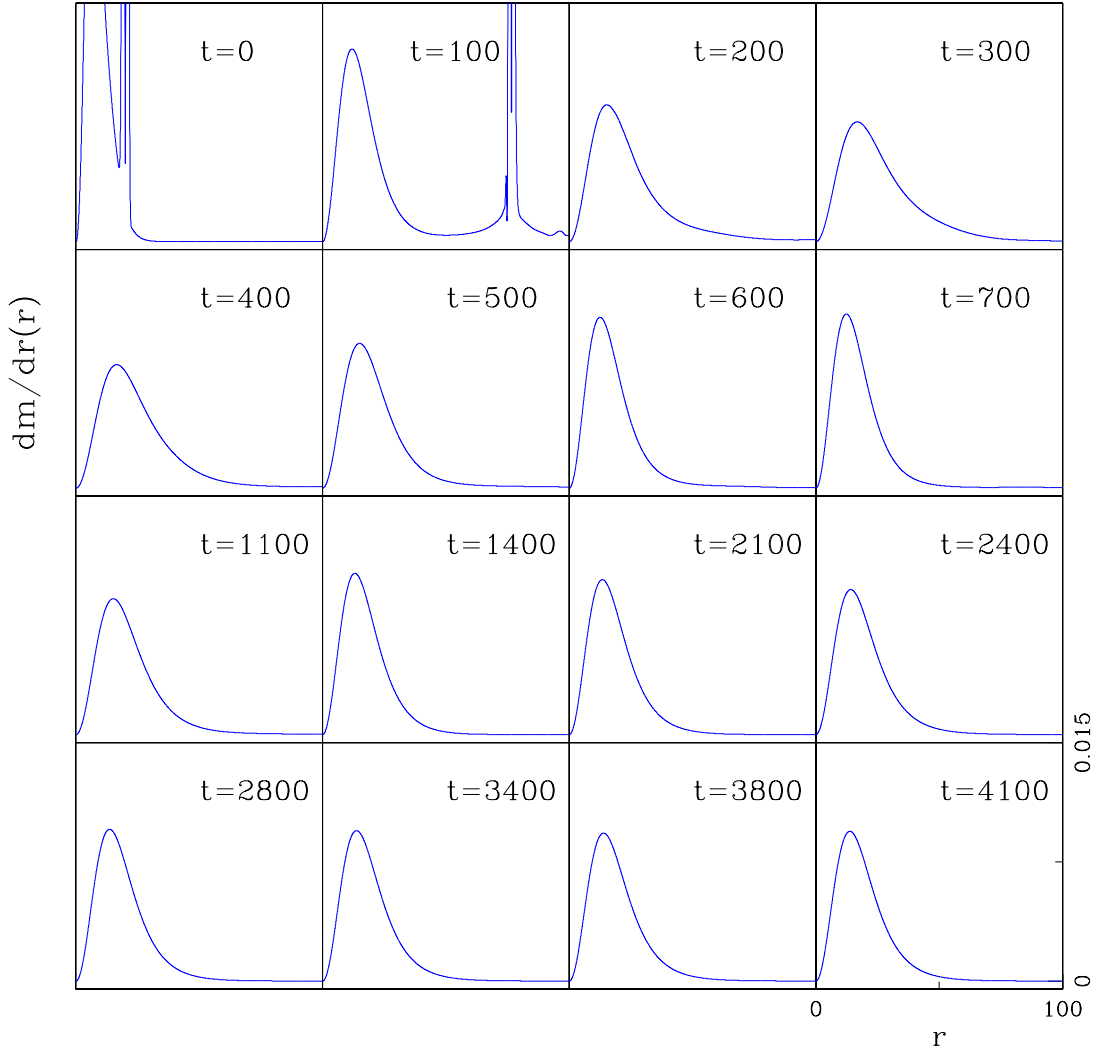}}
\caption[Evolution of a typical ``sphaleron star" solution]
{
This plot shows a sequence of snapshots of a typical ``sphaleron star"
evolution.  The majority of the fields disperse within the first few
frames after which  the regular solution emerges and begins to settle down.
Notice that times 700--1400, 2100--2800, and
3400--4100 show the maximum, median, and minimum values of three different
oscillations.  As can be seen, the first oscillation is quite pronounced,
whereas by the third the amplitude of oscillation is negligible and we are
approaching a static, stable solution.
The solution would appear to be the same to within a few percent for any
values of the initial data that do not produce a black hole.
The mass of the final sphaleron star solution is
approximately $10$--$20\%$ of the mass of the initial configuration.
Note that in setting the scale of the vertical axis, the top portions of the
fields in the first two frames have been cut off.  This has been done to
emphasize the damped oscillations in subsequent frames showing the stable
solution.
For this evolution, we used 10401 mesh points and a Courant factor of 0.5
along with $\eta = 0.1$ and $\lambda = 0.5$.
}
\label{fig:ssshowevolve}
\end{figure}

On varying the center and width of the Yang-Mills potential, 
$\tilde w$, one finds three distinct 
regions of the initial data space.  These correspond, as in \cite{Chhima}, to
generalized Type I collapse, generalized Type II collapse and a ``dispersive"
region in which 
no black hole forms.  On the boundaries between these regions sit appropriate
critical solutions.
It is worth noting, however, that in the region in which no black hole forms,
we no longer observe the complete dispersal of all the matter fields.
Instead, while a majority of the fields do escape to infinity, a nontrivial 
portion of the fields forms a bound state, or ``sphaleron star." 
Shortly after formation, this solution oscillates rapidly, but  
settles down to what appears to be a static solution.  Long evolutions
with $t\sim 30,000$ confirm the stability of this solution.  The mass of this 
stable star is, to within a few percent, 
independent of any of the initial field parameters.  The solution and its 
mass do appear to depend 
on the coupling parameters $\eta$ and $\lambda$ \cite{Maison}. 
Snapshots of a typical evolution in the ``dispersive" regime are shown in 
Figure ~\ref{fig:ssshowevolve}.

\begin{figure}
\epsfxsize=15cm
\centerline{\epsffile{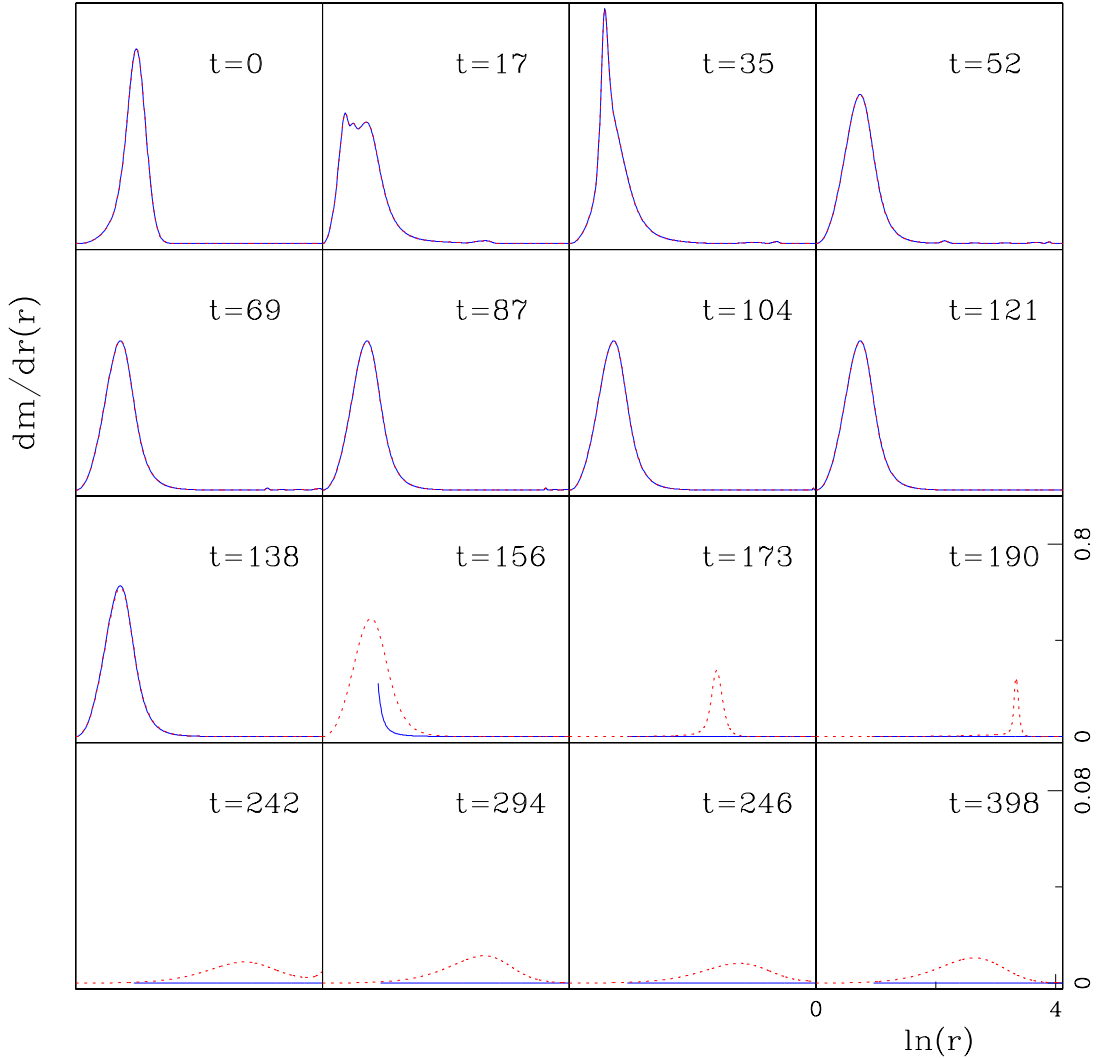}}
\caption[Evolutions on the border between Type I black holes and sphaleron stars]
{
This plot shows two near--critical solutions on the boundary between Type I
collapse and sphaleron star formation.  The critical line between these two
possibilities is characterized by the quasi--static regular solution visible
between $t=52$ and $t=138$.  This solution acts as an attractor for both
near--critical evolutions.  The final state is determined by the initial
data values and is reflected in the evolution away from the attractor.  In
the Type I case (solid line), the majority of the configuration collapses
to form a Schwarschild black hole with a mass gap consistent with Type I
transitions ($t=146$).  In the regular, sphaleron case (dashed line),
about $80$--$90\%$ of
the mass disperses, leaving a stable, bound state with mass independent of
the initial data and location along the critical line.
The final, stable sphaleron star is fundamentally
different from the quasi--static solution that acts as the attractor.  Note
that for visualization purposes, we have rescaled the vertical axis of the
last four frames.  In all frames, the horizontal axis remains unchanged
and measures logarithmic radial coordinate.   In addition, all fields are
plotted so as to be exterior to any horizons.  The gap present in the final
seven frames for the evolution indicated by the solid line is intended to
denote this together with the fact that the final solution on the Type I
side (solid line) is Schwarzschild.
For this evolution, we again used 10401
mesh points and a Courant factor of 0.5 along with $\eta = 0.1$ and
$\lambda = 0.5$.  The width of the Yang--Mills kink is $s=0.8$ and the binary
search was over the center of the kink, $x$. 
}
\label{fig:1sphcrshowevolve}
\end{figure}

In the region with generalized Type I collapse, we note no significant 
change in the dynamics from those similarly exhibited in \cite{Chhima}.  
The collapsing matter forms a black hole with finite mass which, after 
the residual fields have dispersed to infinity, settles down to the 
Schwarzschild solution.  On the Type I critical line separating black 
hole formation and the sphaleron star configuration, we find a regular 
sphaleron as the critical solution analogous to the Bartnik--McKinnon $n=1$
solution.  An example of the critical behavior at this Type I transition 
is given in Figure~\ref{fig:1sphcrshowevolve} in which a sub--critical 
and a super--critical evolution are shown.

\begin{figure}
\epsfxsize=15cm
\centerline{\epsffile{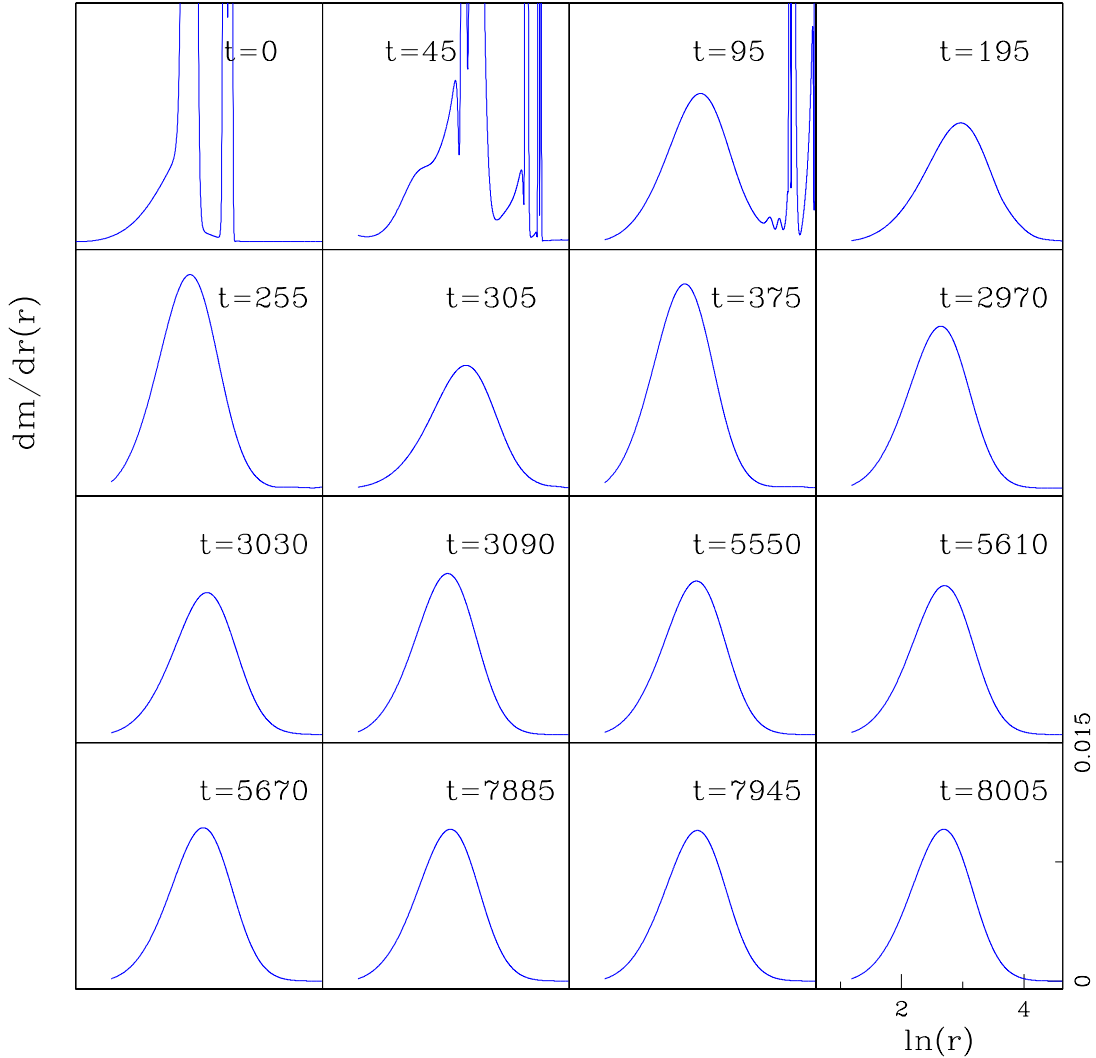}}
\caption[Evolution of a typical hairy black hole solution]
{
This sequence of snapshots shows a typical evolution of a stable, colored
black hole solution within the generalized Type II regime.
The first several frames show the partial dispersal of the initial fields
and the formation of the black hole
solution.
In particular, times 2970--3090, 5550--5670, and 7885--8005 show the maximum,
median, and minimum values of three oscillations.  As can be seen the first
oscillation is quite pronounced, while by the third, the solution is obviously
settling down, ostensibly to a stable, hairy black hole.
As in the previous figure, all fields are plotted so as to be exterior to
any horizons.  The gap in the solution in most of the frames for the evolution
indicated by the solid line is intended to denote
this.  This run was done using 10400 points with a Courant factor of 0.5 and again
with $\eta = 0.1$ and $\lambda = 0.5$.
}
\label{fig:bhshowevolve}
\end{figure}

Within the generalized Type II region, there are some new features.
As in \cite{Chhima}, this region is again characterized by the dynamical
formation of a black hole with finite mass.  As the critical line which
separates dispersion (or strictly, sphaleron star formation) from the black
hole region is approached, the mass of these black holes begins to decrease
such that we interpret the critical transition as Type II.  However, the
black holes that form away from the critical line after the transient
hair has dispersed to infinity do not settle down to Schwarzschild black
holes.  Instead, the final endstate would appear to be a stable, colored
black hole with non--trivial Yang--Mills and Higgs fields outside the
event horizon.  This, of course, is analogous to the sphaleron star that
forms in the no--black--hole region of this system rather than the
complete dispersal of the fields seen in \cite{Chhima}.  The basic
dynamics in this case are illustrated in Figure~\ref{fig:bhshowevolve}.  
The collapsing configuration forms a finite mass black hole
with a significant portion of the remaining field escaping to infinity.
Nevertheless, some of the hair remains behind and within the vicinity of the
event horizon.  This hair oscillates for some time and eventually settles
down to a stable configuration.  Evolutions on the order of $t \sim 30,000$
show no appreciable diminution or instability in the fields.

The mass of the hair in these black hole solutions also seems to be 
independent of the initial data parameters.  Though the radius of the 
black hole will vary with the initial parameters,
the exterior mass remains unchanged.  This observation is consistent with 
and similar to that for the sphaleron stars in which a single stable, regular
solution is found throughout the no--black--hole region.  In addition, like
their regular counterparts, the black hole solutions will depend on the 
parameters $\eta$ and $\lambda$.  Curiously, the mass of this exterior hair 
is very nearly the same value as the mass of the sphaleron star.  
Thus, in one sense, these hairy black hole 
solutions can be thought of as sphaleron star solutions within which the 
central density increases to the point that a horizon forms.  This is 
similar in turn to gravitating t'Hooft-Polyakov monopoles.  For certain 
values of the coupling, the monopoles can have a black hole form at their 
center.
 
For the solutions near the threshold separating black hole formation and 
dispersion, we find 
hints that these are indeed Type II critical solutions and that the black 
hole mass scales as expected.  However, we stress again that our unigrid code
is not able to settle this issue definitively and that it awaits additional
study.   

\begin{figure}
\epsfxsize=15cm
\centerline{\epsffile{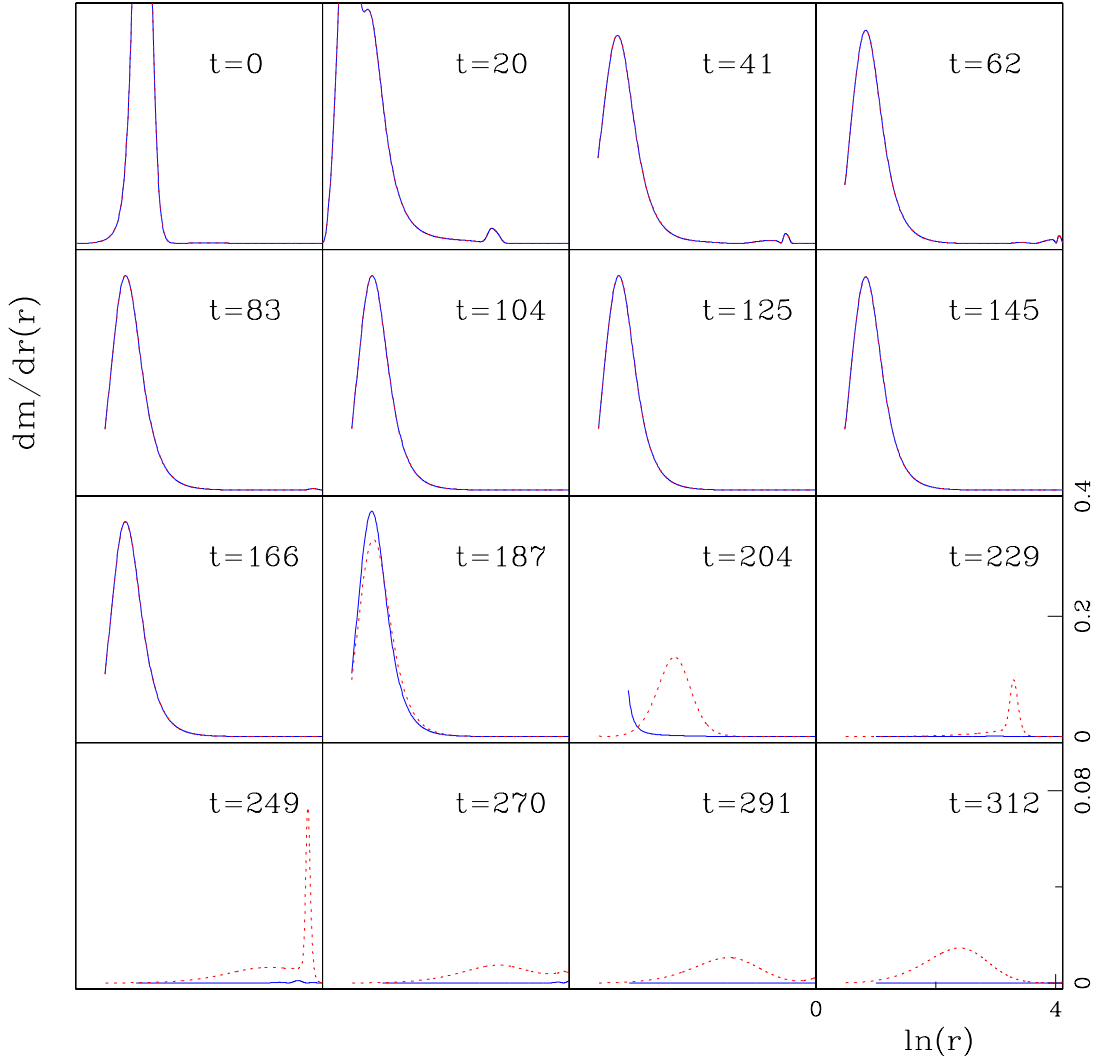}}
\caption[Evolutions on the border between Type I and Type II black holes]
{This plot shows two near-critical evolutions on the border between generalized
Type I and Type II black hole formation.  This critical line between these
two possibilities is characterized by a family of black holes parameterized
by their horizon radius and possessing non-trivial Yang--Mills--Higgs hair
outside the horizon.  These colored black holes serve as
the attractors for these two types of collapse and can be seen here between
$t=41$ and $t=166$.
The final state of the collapse can be distinguished
by the subsequent evolution of the fields away from the attractor.  On
the Type I side (the solid line), the hair falls into the horizon (at
$t\approx204$), adding to the mass of the black hole, and resulting finally
in a Schwarschild black hole.  On the Type II side (the dashed line),
the majority of the hair disperses to infinity.  However, between
$10$--$20\%$ of the initial
mass of the system (depending on where one is along the critical line)
remains behind, eventually settling down and forming stable, Yang-Mills-Higgs
hair outside a black hole.
Note that the horizontal axis is the natural
logarithm of the radial coordinate and that in the final four frames, the
vertical axis is rescaled to better visualize the remaining hair in the
Type II case.  Again, all fields have been plotted so as to be exterior to
any horizons.  The gap(s) present in the plotted solutions is intended to
denote this.  This evolution used 10401 mesh points and a Courant factor
of 0.5 and again with $\eta = 0.1$ and $\lambda = 0.5$.
The width of the Yang--Mills kink is $s=0.7$ and the binary
search was over the center of the kink, $x$. 
}
\label{fig:12crshowevolve}
\end{figure}

\begin{figure}
\epsfxsize=15cm
\centerline{\epsffile{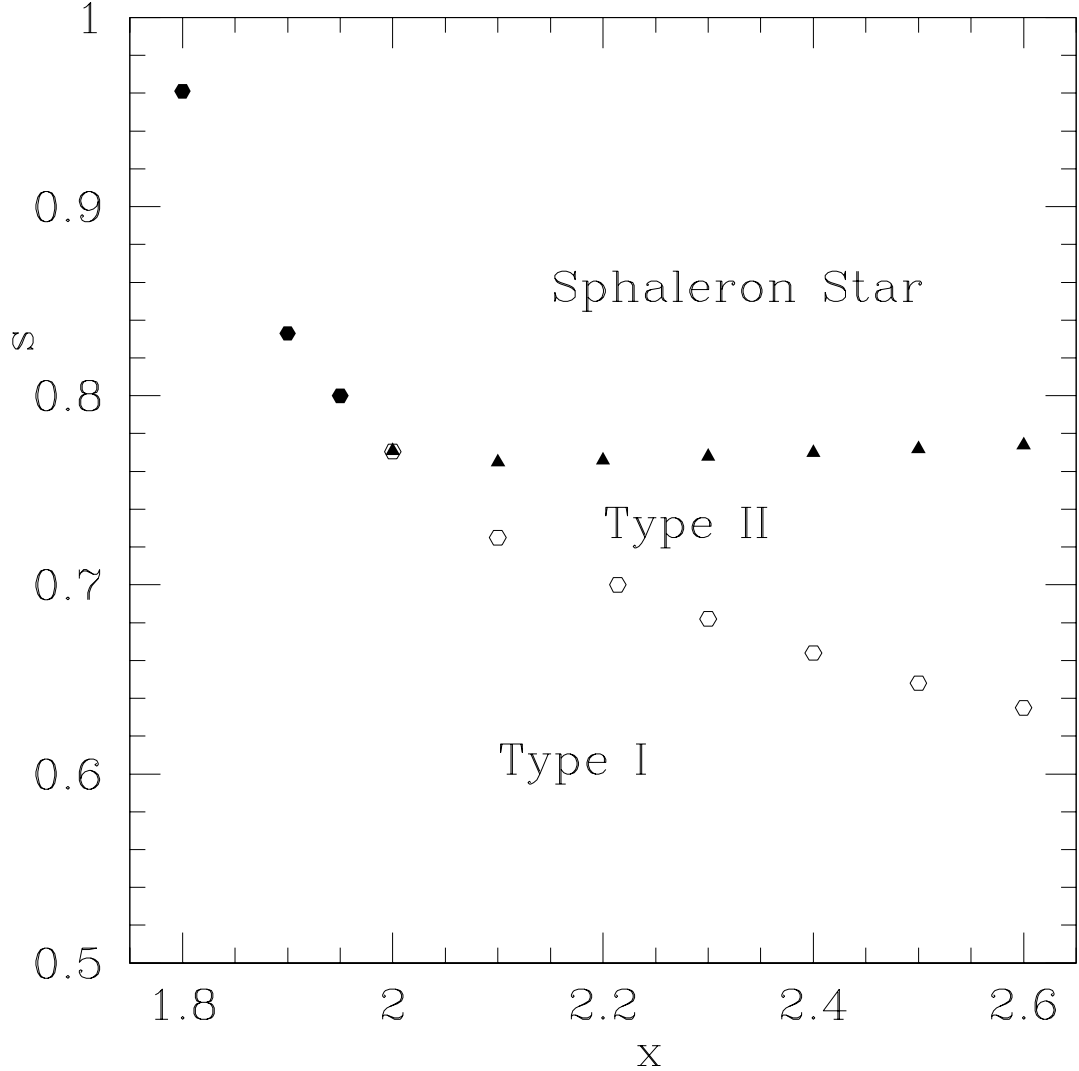}}
\caption[Phase space of ``bland" Higgs configuration]
{
This plot is of the initial data space and illustrates the endstates of
collapse as the width, $s$, and center, $x$, of the initial Yang--Mills field
are varied.
The filled hexagons give the boundary between the formation of sphaleron
star solutions and Type I black holes.  The filled triangles represent the
boundary between the formation of sphaleron stars and Type II hairy
black holes.  The open hexagons depict the colored critical solutions on the
boundary between Type I and Type II black holes.  Near these critical
solutions, type is depicted not only by the existence of the stable hair,
but also by the transient hair either dispersing
or falling down the horizon.   Note the similarity between this and Figure 4 of
\cite{Chhima}.  Each point depicted on the plot represents an evolution in
which $|p-p^{*}| < 10^{-5}$.  For each evolution we again used 10401 mesh
points and a Courant factor of 0.5
along with $\eta = 0.1$ and $\lambda = 0.5$.
}
\label{fig:bland}
\end{figure}

We are able, though, to consider the transition between the different types 
of dynamical collapse in the black hole regime.  Again, we find a family of 
critical solutions separating generalized Type I collapse from 
generalized Type II collapse.  These critical solutions are themselves 
sphaleron black holes parameterized by their horizon radius such that as one
moves away from the ``triple point" in Figure~\ref{fig:bland}, that radius 
increases.  On the Type II side of this line, near--critical evolutions 
have dynamics described above with the collapsing configuration forming a 
black hole of finite mass with non--trivial hair outside.  However, as the 
transition between Type II and Type I is approached, there is an intermediate 
solution -- a hairy black hole -- which forms and to which the evolving solution is attracted.
This intermediate solution is unstable and eventually collapses.  For initial 
configurations on the Type II side, the collapse is distinctive in that very
little  
of the exterior fields falls into the black hole.  Rather, some of it is 
dispersed to infinity while the remainder reconstitutes in a new and different 
colored configuration outside the black hole already present.   
This is shown in Figure~\ref{fig:bhshowevolve} for a generic collapse in the 
Type II regime as well as in the last frames of 
Figure~\ref{fig:12crshowevolve} for a near--critical evolution.

On the Type I side of the critical line, similar near--critical evolutions
exhibit the same early time dynamics with the formation of a finite mass 
black hole and the approach to the intermediate hairy black hole.  However, 
as the critical line is approached, this unstable black hole now collapses and 
loses most of its hair into the black hole causing it to grow in 
size.  A picture of both collapse dynamics is shown in 
Figure~\ref{fig:12crshowevolve}. 

So far, our entire description has been within the context of varying two of 
the initial data parameters that describe the Yang--Mills field.  We have 
chosen to vary the center and width of the Yang--Mills kink, but the same 
results hold for other Yang--Mills parameters.  However, there does appear 
to be a non--trivial difference in our description of the initial data space if
we include a parameter of the Higgs field in the set of two parameters which
we vary in order to search for critical behavior.   

In the original, so-called ``bland" Higgs case, the structure of the 
phase space is very similar to that found in \cite{Chhima}, namely Type I 
and II critical collapse separated by the previously discussed colored, 
critical solution.  This is shown in Figure~\ref{fig:bland}. 

\begin{figure}
\epsfxsize=15cm
\centerline{\epsffile{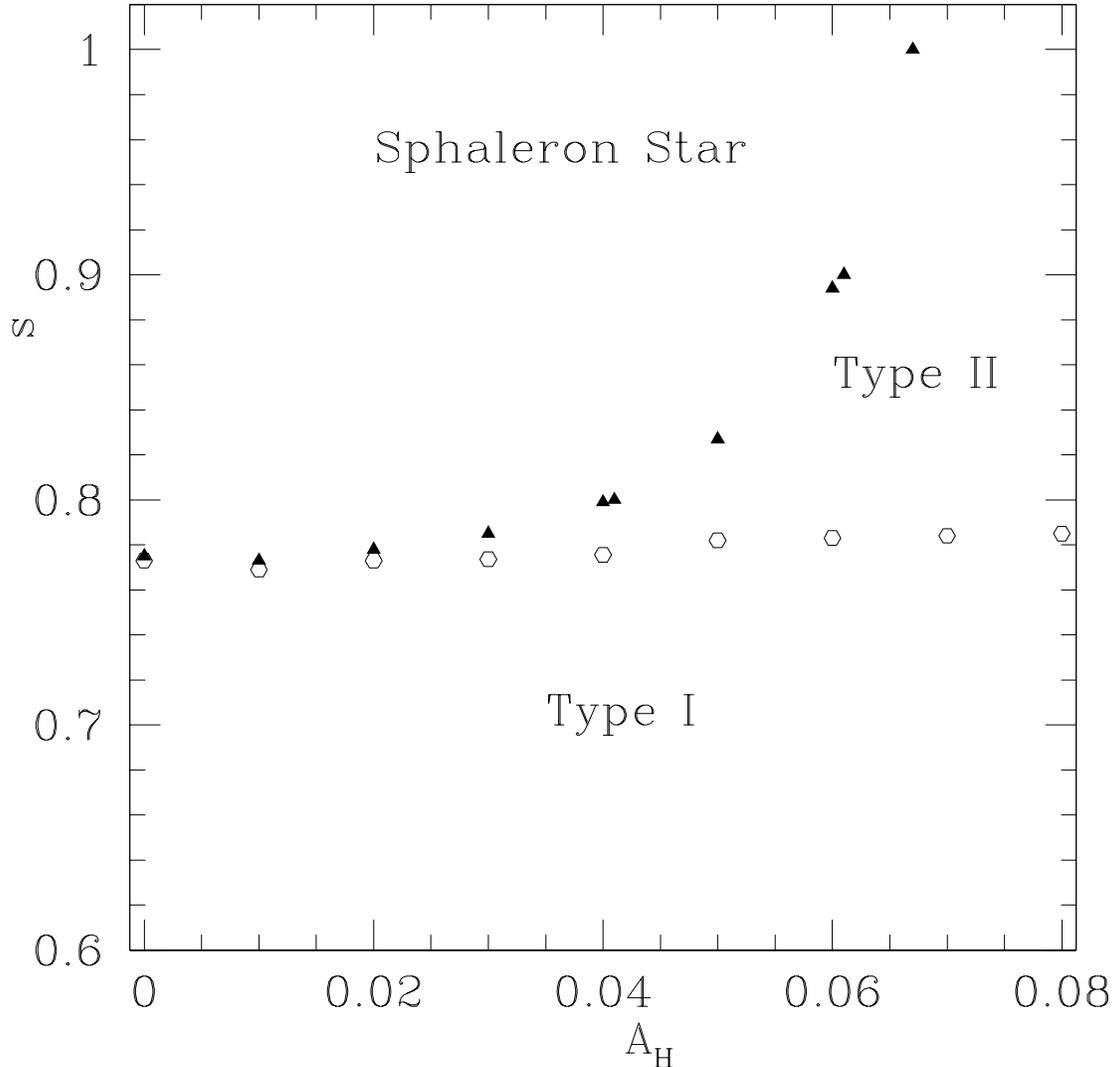}}
\caption[Phase space of ``pulse" Higgs configuration]
{
This is a second plot of the initial data space and illustrates the endstates
of collapse as the width, $s$, of the initial Yang-Mills field and
the amplitude, $A_H$, of of the Higgs pulse are varied.  It should be
noted that the structure for negative values of $A_H$ is symmetric across
the $s$-axis.  Again, the filled triangles
represent the boundary between the sphaleron stars and the Type II hairy
blackholes while the open hexagons denote the location of the critical
solutions separating generalized Type I and Type II collapse.  Note that
the overlap of some of the polygons denotes only that the boundaries are
near one another, not that they overlap.  In this mixed phase space, we
find no region in which Type I behavior borders the region of sphaleron star
formation.  As a result, it would appear that this portion of the initial
data  space does not contain a triple point.  Note the similarity in the
limits of the vertical axis in this and the previous figure.
Each point on the respective critical lines represents a critical solution
at a level $|p-p^{*}|/p^{*} < 10^{-5}$.  For each evolution we again used
10401 mesh points and a Courant factor of 0.5 along with $\eta = 0.1$
and $\lambda = 0.5$.
}
\label{fig:pulse}
\end{figure}

However, in the event that we turn on the exponential, perturbing pulse 
for the Higgs field, and look at the phase space including one Yang--Mills 
parameter and one Higgs parameter, we see a different picture, shown in 
Figure~\ref{fig:pulse}.  In this case and for the region of the parameter 
space explored, there appears to be no critical boundary between Type I 
collapse and the formation of a sphaleron star.  Instead, Type II collapse 
would seem to border the region in which the regular solution is formed.  
It would therefore seem that in this region of the initial data space, 
there is no ``triple point."  One might have imagined that as one varied the 
amplitude of the exponential portion of the Higgs pulse that there would be 
such a point in each $xs$ plane for a given $A_H$.  However, this does not 
seem to be the case and is an issue deserving of more investigation.  

\begin{figure}
\epsfxsize=7.5cm
\centerline{\epsffile{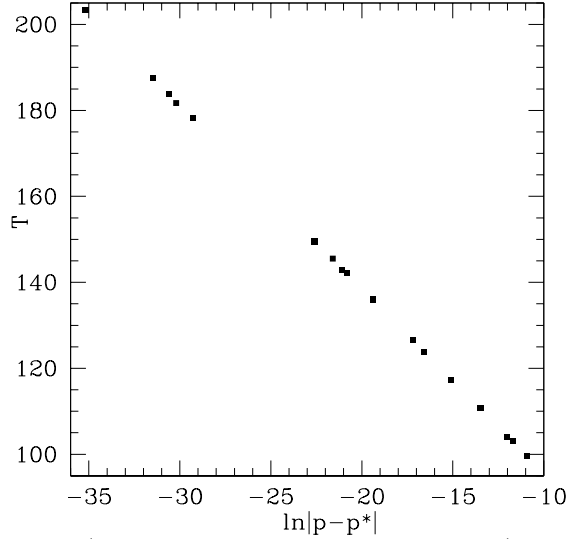}}
\caption[Scaling of near critical solutions]
{
This plot gives the elapsed time (as measured by an observer at infinity)
spent by an evolving configuration in the sphaleron regime as the critical line
separating generalized Type I collapse and sphaleron star formation is
approached.  In Type I collapse, there is a scaling relation for the time given
by $T = -\lambda_{\rm I} |p-p^{*}|$ where $\lambda_{\rm I}$ is the inverse
Lyapounov exponent of the corresponding unstable mode of the critical
solution sitting on the threshold between black hole formation and dispersion.
In the current case, we find from a least squares fit that
$\lambda_{\rm I}=4.27(1)$.  We define the elapsed time, $T$, as the time
from the beginning of the evolution until the pulse crosses $r=40$.
As before, for these evolutions, we used 10401 mesh points and a Courant
factor of 0.5 along with $\eta = 0.1$ and $\lambda = 0.5$.
}
\label{fig:scale1ss}
\end{figure}

\begin{figure}
\epsfxsize=7.5cm
\centerline{\epsffile{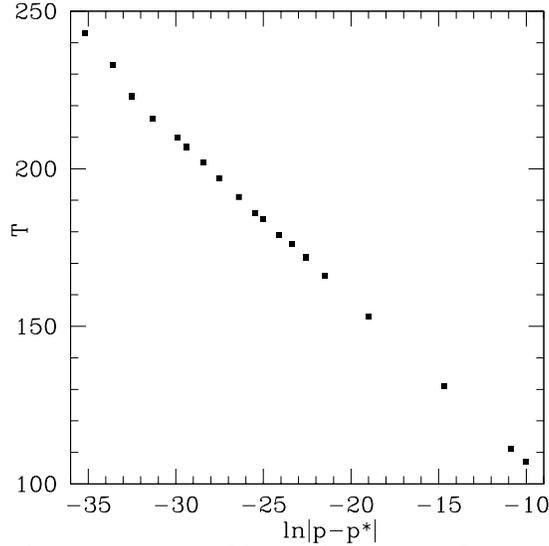}}
\caption[Scaling of near critical solutions]
{
This graph gives the elapsed time (as measured by an observer at infinity)
spent by an evolving configuration in the Type II regime as the critical line
separating generalized Type I and Type II collapse is approached.
As the critical solution is approached, the evolution spends more and more time
on the critical solution and we expect a linear relationship between that
time and logarithmic distance in the initial data space: 
$T = -\lambda_{\rm I--II} |p-p^{*}|$.  The slope of this line, 
$\lambda_{\rm I--II}$, is the inverse Lyapounov exponent of the corresponding 
unstable mode of the critical solution.  From a least square fit, we find 
$\lambda = 5.27(1)$.  The time on critical is defined as the time
between the beginning of the evolution until the pulse in the type II case
crosses $r=40$.
As before, for these evolutions, we used 10401 mesh points and a Courant
factor of 0.5 along with $\eta = 0.1$ and $\lambda = 0.5$.
}
\label{fig:scale12}
\end{figure}

Finally, we note that the critical solutions separating the two types of
collapse and Type I collapse from sphaleron star formation exhibit time 
scaling as would be expected.  As the single, unstable mode characteristic 
of each critical solution is tuned out, near--critical solutions spend 
increasing amounts of time as measured by an asymptotic observer on the 
critical solution.  These scaling relations are given by 
$T \approx -\lambda \ln |p-p^*|$ where $\lambda$ is the characteristic 
time scale for the collapse of the unstable critical solution.  It 
corresponds to the inverse Lyapounov exponent of the unstable mode. 
Such scaling relations specific to points on the relevant critical lines
are shown in Figures~\ref{fig:scale12} and \ref{fig:scale1ss}.

\section{Discussion}
\label{sec:discussion}
We have presented evidence for critical phenomena in the gravitational collapse 
of $SU(2)$ sphaleron configurations of Yang--Mills--Higgs fields.  In many
respects, this collapse is qualitatively similar to 
that in the Einstein--Yang--Mills system 
but does have some notable suprises.  The critical behavior is seen again in 
three possible transitions.  On each of these transitions sit critical 
solutions which serve as intermediate attractors for nearby evolutions in the 
initial data space.  Near the critical line separating Type I collapse from the
regular solution as well as for the critical line separating the two types of dynamical black 
hole formation, there are time scaling relations as the near--critical 
solutions approach the critical solutions.  In addition, the mass of 
the black holes formed in the appropriate region will exhibit a mass gap in
crossing these critical lines.  Near the critical line separating Type II collapse from the regular endstate region, we have indications that the mass of the 
black hole scales without a mass gap, but again, due to our unigrid code, we
can not settle this conclusively although expectations and indications would
bear this out.  

Among the surprises in this model are that in certain regions
of the initial data space, we find that 
regular, stable, sphaleron solutions are produced rather than the purely
dispersive regime seen in \cite{Chhima}. 
However, we also
have stable, hairy or colored black holes produced in the supercritical, black
hole regime.  This contrasts again with earlier results in which the endstate
in spherical symmetry was always a Schwarzschild black hole.  Thus, the 
transition here between generalized Type I and Type II collapse is a transition
between {\em different} types of black holes.  Within the Type I region, the 
endstate is always a Schwarzschild black hole with the exterior gauge and Higgs
fields either falling into the existing black hole or dispersing to infinity. 
In the Type II region, the final black holes are stable, colored black holes.  
It is worth reemphasizing that the colored black hole solutions sitting on the 
critical line separating types of black hole collapse are not the same as
the stable, colored black holes that are the final endstates in the Type II
region.  This can be seen most easily in Figure~\ref{fig:12crshowevolve}. 
 
It is also noteworthy that the existence of all the solutions which we find is contingent
on the magnetic ansatz within which we have chosen to work.  In general, both the regular and 
colored black hole solutions which we find to be stable endstates of collapse
are expected to be unstable based on a linear perturbation analysis \cite{Bbmsv}.  
However, such an analysis assumes that both the gravitational and sphaleron 
sectors in the theory are perturbed.  Our evolutions perturb only the 
gravitational sector.  It is reasonable to assume that the stable solutions
which we find will become unstable on perturbation of the Yang--Mills gauge 
field away from the magnetic ansatz.  We hope to address this issue in 
future work.   

Another issue for future consideration is the structure of the initial data
space.  A curiosity of our current results is that the ``triple point" would
seem to disappear as the Higgs field is varied.  As a result, the boundary 
between regular endstates (sphaleron star formation here) and black hole 
formation is taken up entirely by a Type II transition.   

Nonetheless, given the structure of the initial data space, one can draw an
analogy with the gravitating monopole case in which a small black hole can 
form within a t'Hooft-Polyakov monopole coupled to gravity.  
This stable object can be rendered 
unstable above a maximum value of the horizon radius at which point the 
exterior Yang-Mills-Higgs hair will either fall into the black hole or disperse
leaving a final Schwarzschild black hole.  A similar thing happens in the 
current sphaleron case.  For example, following a line of constant $x$ in 
Figure~\ref{fig:bland} that intersects each region we see that as $s$ decreases, 
one can interpret the process in a similar way.  A regular solution develops
a small, stable black hole at the center which (with decreasing width of the initial
Yang-Mills potential, $s$) increases 
in size until the combined sphaleron and black hole system becomes unstable 
and is replaced with a larger Schwarzschild black hole.  
As a result, it would be interesting to consider the full dynamical 
evolution of the gravitating monopole and compare with the sphaleron case
reported here.

\section*{Acknowledgements}
\label{sec:ack}
This research has been supported in part by NSF grants
PHY-9900644 and PHY-0139782.  EWH would like 
to thank A.~Wang for useful discussions as well as M.~Choptuik     
and R.~Marsa for their work on an earlier
code which served as the precursor to the one described here.



\vfil\eject

\vfil\eject

\vfil\eject

\vfil\eject

\vfil\eject

\vfil\eject

\end{document}